\documentstyle[amssymb,epsf,cite,11pt]{article}

\newcommand{\be}{\begin{equation}}
\newcommand{\ee}{\end{equation}}

\begin{document}

\title{\bf The subordinated processes controlled by a family of subordinators
and corresponding Fokker-Planck type equations}
\author{Long Shi$^{1,2}$, Zu-Guo Yu$^{1,3}$\thanks{Corresponding to yuzuguo@aliyun.com}, Hai-Lan Huang$^1$, Zhi Mao$^1$ and Ai-Guo Xiao$^1$\\
{\small$^1$ Hunan Key Laboratory for Computation and Simulation in
Science and Engineering and }\\
{\small Key Laboratory of Intelligent Computing and Information
Processing of Ministry of Education,}\\
{\small Xiangtan University, Xiangtan,  Hunan 411105, China.}\\
{\small $^2$ School of Science, Central South University of Forest
and Technology,}\\
{\small  Changsha, Hunan 410004, China.}\\
{\small $^{3}$School of Mathematical Sciences, Queensland University of Technology,}\\
{\small GPO Box 2434, Brisbane, Q4001, Australia.}
 }
\date{}
\maketitle

\begin{abstract}
In this work, we consider subordinated processes controlled by a
family of subordinators which consist of a power function of time
variable and a negative power function of $\alpha-$stable random
variable. The effect of parameters in the subordinators on the
subordinated process is discussed. By suitable variable
substitutions and Laplace transform technique, the corresponding
fractional Fokker-Planck-type equations are derived. We also
compute their mean square displacements in a free force field. By
choosing suitable ranges of parameters, the resulting subordinated
processes may be subdiffusive, normal diffusive or superdiffusive.
\end{abstract}

{\bf Key words}: subordinated process; subordinator;
Fokker-Planck-type equation; mean square displacement; variable
substitution; Laplace transform

\section{Introduction}

\ \ \ \ Anomalous diffusion is one of the most universal phenomena
in nature [1-3]. In one dimension, it is characterized by a mean
square displacement (MSD) of the form
\begin{equation}
\label{(1)} \langle (\Delta x)^2 \rangle \sim Dt^{\alpha}
\end{equation}
with $\alpha\neq 1$, deviating from the linear dependence on time
found for normal diffusion [1]. The coefficient $D$ is generalized
diffusion constant. It is called subdiffusion for $0<\alpha<1$ and
superdiffusion for $\alpha>1$ [4].

One of the simplest models to describe anomalous diffusion is
continuous time random walk (CTRW), introduced by Montroll and
Weiss [5]. Let $W(x,t)$ be the probability density function (PDF)
of finding a particle (with continuous distribution of step
length) at point $x$ at time $t$, one has [5]
\begin{equation}
\label{(2)} W(x,t)=\sum_{n}f(x,n)g_{n}(t),
\end{equation}
where $f(x,n)$ is the PDF to find the particle at point $x$ after
$n$ steps and $g_{n}(t)$ is the probability to make exactly $n$
steps up to time $t$. Here, the random number of steps $n$ plays
the role of operational time governing the evolution of the
system.

A continuous realization of Eq.(2) is
\begin{equation}
\label{(3)} W(x,t)=\int_{0}^{\infty}f(x,\tau)g(\tau,t)d\tau,
\end{equation}
which has been considered by Fogedby in [6] in terms of a coupled
system of It\^{o} stochastic differential equations for position
$x$ and time $t$
\begin{equation}
\label{(4)} \frac{dx(\tau)}{d\tau}=F(x)+\Gamma(\tau), \hspace{1cm}
\frac{dt(\tau)}{d\tau}=\eta(\tau),
\end{equation}
where $F(x)$ is an external force, $\Gamma(\tau)$ and $\eta(\tau)$
are random noise sources that are assumed to be independent. In
this frame, the position $x$ and the physical time $t$ are
parameterized in terms of the continuous variable $\tau$ which may
be considered as operational time. The random walk in physical
time $Y(t)$ is then given by the combined process $Y(t)=x(s(t))$,
where $\tau=s(t)$ is the inverse process to $t(\tau)$ defined as
\begin{equation}
\label{(5)} s(t)=inf\{\tau>0: t(\tau)>t\},
\end{equation}
a continuous analogue of the random number of the steps $n(t)$.

In formula (3) $W(x,t)$, $f(x,\tau)$ and $g(\tau,t)$ are the PDFs
of the processes $Y(t)$, $x(\tau)$ and $s(t)$, respectively. From
the opinion of subordination, which is originated by Bochner [7],
the combined process $Y(t)$ can be said to be subordinated to
parent process $x(\tau)$ using the operational time $\tau=s(t)$.
The process $s(t)$ is also called subordinator. It is worth to
note that the continuous time random processes in phase
(position-velocity) space based on the subordination were
discussed in Refs. [8-10].

In system (4), the most popular choice on random noise sources
$\Gamma(\tau), \eta(\tau)$ is taken $\Gamma(\tau)$ as a Gaussian
noise and $\eta(\tau)$ as a fully skew $\alpha-$stable L\'{e}vy
noise (i.e. $s(t)$ is an inverse $\alpha-$stable subordinator). In
recent years, the system of coupled Langevin equations (4) with
inverse $\alpha-$stable subordinator is becoming a hot topic
[11-25]. The inverse tempered $\alpha-$stable subordinator and the
infinitely divisible inverse subordinators were also considered
[26-34].

However, one cannot obtain directly characteristics of
subordinator $s(t)$ from the definition (5). In [35], Meerschaert
and Scheffler showed that the $\alpha-$self-similar process $E(t)$
has the same distribution as the process $t^{\alpha}Z^{-\alpha}$,
where $Z$ is an $\alpha-$stable random variable, i.e. $\langle
e^{-uZ}\rangle=e^{-u^{\alpha}}$. Motivated by this, we assume that
the subordinator $s(t)$ has the following three forms
$s(t)=t^{\alpha}Z^{-\alpha}$, $s(t)=t^{\beta}Z^{-\alpha}$ or
$s(t)=t^{\beta}Z^{-\beta}$ and then discuss the resulting
subordinated processes $x(s(t))$.

This paper is organized as follows. In section 2, we introduce the
subordinated process controlled by subordinator
$s(t)=t^{\alpha}Z^{-\alpha} (0<\alpha<1)$ and connect the
subordinated process $x(s(t))$ with a time-fractional
Fokker-Planck equation (FPE). In section 3, we discuss the
subordinated process $x(s(t))$ with $s(t)=t^{\beta}Z^{-\alpha}
(0<\alpha<1, 0<\beta<2)$. The subordinated process $x(s(t))$ with
$s(t)=t^{\beta}Z^{-\beta} (0<\beta<2)$ is considered in section 4.
The conclusions are presented in section 5.

\section{The subordinated process with subordinator $s(t)=t^{\alpha}Z^{-\alpha}$}

\ \ \ \ Assumed that the \textit{parent process} $x(\tau)$
satisfies the Langevin equation
\begin{equation}
\label{(6)} \frac{dx(\tau)}{d\tau}=F(x(\tau))+\Gamma(\tau),
\end{equation}
where $\Gamma(\tau)$ is a white Gaussian noise with $\langle
\Gamma(\tau)\rangle=0, \langle
\Gamma(\tau)\Gamma(\tau')\rangle=2\delta(\tau-\tau')$. Thus, the
PDF $f(x,\tau)$ obeys the following ordinary FPE [36]:
\begin{equation}
\label{(7)} \frac{\partial}{\partial \tau}f(x,\tau)=
(-\frac{\partial}{\partial x}F(x)+ \frac{\partial^2}{\partial
x^2})f(x,\tau)\triangleq L_{FP}f(x,\tau).
\end{equation}

Now we assume $s(t)=t^{\alpha}Z^{-\alpha} (0<\alpha<1)$, where $Z$
is an $\alpha-$stable random variable with the PDF
$\psi_{\alpha}(t)$, i.e. the Laplace transform $ \langle
e^{-uZ}\rangle=\int_{0}^{\infty}\psi_{\alpha}(t)e^{-ut}dt=e^{-u^{\alpha}}$.

Since
\begin{equation}
\label{(8)} P(s(t)\leq \tau) = P(Z\geq \frac{t}{\tau^{1/\alpha}})
=\int_{\frac{t}{\tau^{1/\alpha}}}^{\infty}\psi_{\alpha}(t')dt',
\end{equation}
we have the PDF $g(\tau,t)$ of subordinator $s(t)$ as
\begin{equation}
\label{(9)} g(\tau,t)=
\frac{1}{\alpha}\frac{t}{\tau^{1+1/\alpha}}\psi_{\alpha}(\frac{t}{\tau^{1/\alpha}}).
\end{equation}

Recalling the following Laplace transform formulas
\begin{equation}
\label{(10)} \int_{0}^{\infty}\varphi(at)e^{-ut}dt=
\frac{1}{a}\int_{0}^{\infty}\varphi(t')e^{-\frac{u}{a}t'}dt'
=\frac{1}{a}\widetilde{\varphi}(\frac{u}{a}), \hspace{0.5cm}a>0,
\end{equation}
and
\begin{equation}
\label{(11)} \int_{0}^{\infty}t\varphi(t)e^{-ut}dt
=-\frac{d}{du}\int_{0}^{\infty}\varphi(t)e^{-ut}dt =
-\frac{d}{du}\widetilde{\varphi}(u),
\end{equation}
we get the Laplace transform of $g(\tau,t)$ as
\begin{equation}
\label{(12)}
\widetilde{g}(\tau,u)=u^{\alpha-1}\exp\{-u^{\alpha}\tau\}.
\end{equation}

Substituting $\widetilde{g}(\tau,u)$ into Eq.(3), we get the PDF
$W(x,t)$ of the subordinated process $Y(t)=x(s(t))$ in Laplace
space:
\begin{equation}
\label{(13)}
\begin{array}{lll}
\widetilde{W}(x,u)
&=& \int_{0}^{\infty}f(x,\tau)\widetilde{g}(\tau,u)d\tau \\
\\
&=& \int_{0}^{\infty}f(x,\tau)u^{\alpha-1}\exp\{-u^{\alpha}\tau\}d\tau\\
\\
&=& u^{\alpha-1}\widetilde{f}(x,u^{\alpha}).
\end{array}
\end{equation}

Since $f(x,\tau)$ satisfies Eq.(7), one get in Laplace space
\begin{equation}
\label{(14)} u\widetilde{f}(x,u)-f(x,0)=L_{FP}\widetilde{f}(x,u).
\end{equation}

Assumed that $W(x,0)=f(x,0)=\delta(x)$, from Eq.(13) and Eq.(14),
we have
\begin{equation}
\label{(15)}
\begin{array}{lll}
u\widetilde{W}(x,u)-W(x,0)
&=& u^{\alpha}\widetilde{f}(x,u^{\alpha})-f(x,0)\\
\\
&=& L_{FP}\widetilde{f}(x,u^{\alpha})\\
\\
&=& u^{1-\alpha}L_{FP}\widetilde{W}(x,u).
\end{array}
\end{equation}

Taking Laplace inverse transform for $u\rightarrow t$ in Eq.(15),
we obtain the following time-fractional FPE [4,37-39]
\begin{equation}
\label{(16)} \frac{\partial W(x,t)}{\partial
t}=_{0}D_{t}^{1-\alpha} L_{FP}W(x,t),\hspace{0.5cm}0<\alpha<1,
\end{equation}
where $_{0}D_{t}^{1-\alpha}$ denotes the Riemann-Liouville
fractional derivative [40], defined by
\begin{equation}
\label{(17)} _{0}D_{t}^{1-\alpha}f(t)=
\frac{1}{\Gamma(\alpha)}\frac{\partial}{\partial
t}\int_{0}^{t}(t-s)^{\alpha-1}f(s)ds.
\end{equation}
Eq.(16) can be also derived from a generalised master equation
[41], or from continuous time random walks (CTRWs) [42].

We are also interested in the MSD of the subordinated process
$Y(t)$ in free force field. One can compute $\langle
x^2\rangle(t)$ easily in terms of Eq.(16) with natural boundary
conditions:
\begin{equation}
\label{(18)} \langle
x^2\rangle(t)=\int_{-\infty}^{\infty}x^{2}W(x,t)dx
=\frac{2}{\Gamma(\alpha+1)}t^{\alpha}, \hspace{0.5cm}0<\alpha<1.
\end{equation}
From (18), we conclude that the proposed subordinated process is
subdiffusive.

\section{The subordinated process with subordinator $s(t)=t^{\beta}Z^{-\alpha}$}

\ \ \ \ In this section, we assume that $x(\tau)$ satisfies
Langevin equation (6) and $s(t)$ has the form
$s(t)=t^{\beta}Z^{-\alpha} (0<\alpha<1, 0<\beta<2)$.

Since
\begin{equation}
\label{(19)} P(s(t)\leq \tau) =P(Z\geq
\frac{t^{\beta/\alpha}}{\tau^{1/\alpha}})
=\int_{\frac{t^{\beta/\alpha}}{\tau^{1/\alpha}}}^{\infty}\psi_{\alpha}(t')dt',
\end{equation}
we have
\begin{equation}
\label{(20)} g(\tau,t)=
\frac{1}{\alpha}\frac{t^{\beta/\alpha}}{\tau^{1+1/\alpha}}
\psi_{\alpha}(\frac{t^{\beta/\alpha}}{\tau^{1/\alpha}}).
\end{equation}

Setting
\begin{equation}
\label{(21)} h(\tau,t)=
\frac{1}{\alpha}\frac{t}{\tau^{1+1/\alpha}}\psi_{\alpha}(\frac{t}{\tau^{1/\alpha}}),
\end{equation}
one can easily obtain
\begin{equation}
\label{(22)} g(\tau,t)=h(\tau,t_1)
\end{equation}
with $t_1=t^{\beta/\alpha}$.

Since
\begin{equation}
\label{(23)} W(x,t)=\int_{0}^{\infty}f(x,\tau)g(\tau,t)d\tau,
\end{equation}
we have
\begin{equation}
\label{(24)}
\begin{array}{lll}
\frac{\partial}{\partial t}W(x,t)
&=& \int_{0}^{\infty}f(x,\tau)\frac{\partial}{\partial t}g(\tau,t)d\tau \\
\\
&=& \int_{0}^{\infty}f(x,\tau)\frac{\partial h(\tau,t_1)}{\partial t_1}\frac{dt_1}{dt}d\tau \\
\\
&=& \frac{\partial}{\partial
t_1}\int_{0}^{\infty}f(x,\tau)h(\tau,t_{1})d\tau\frac{dt_1}{dt}.
\end{array}
\end{equation}
Setting $W_{1}(x,t)=\int_{0}^{\infty}f(x,\tau)h(\tau,t)d\tau$, by
comparing Eq.(21) with Eq.(9), we obtain that $W_{1}(x,t)$
satisfies
\begin{equation}
\label{(25)} \frac{\partial W_{1}(x,t)}{\partial t}
=_{0}D_{t}^{1-\alpha} L_{FP}W_{1}(x,t).
\end{equation}

Substituting Eq.(25) into Eq.(24), we have
\begin{equation}
\label{(26)}
\begin{array}{lll}
\frac{\partial}{\partial t}W(x,t)
&=& \frac{\partial}{\partial t_1}W_{1}(x,t_1)\cdot\frac{dt_1}{dt}\\
\\
&=& \frac{1}{\Gamma(\alpha)}\frac{\partial}{\partial t_1}
\int_{0}^{t_1}(t_1-s)^{\alpha-1}L_{FP}W_{1}(x,s)ds\cdot\frac{dt_1}{dt}\\
\\
&=&  \frac{1}{\Gamma(\alpha)}\frac{\partial}{\partial t}
\int_{0}^{t_1}(t_1-s)^{\alpha-1}L_{FP}W_{1}(x,s)ds\\
\\
&=&  \frac{1}{\Gamma(\alpha)}\frac{\partial}{\partial t}
\int_{0}^{t^{\beta/\alpha}}(t^{\beta/\alpha}-s)^{\alpha-1}L_{FP}W_{1}(x,s)ds.
\end{array}
\end{equation}

After replacing the variable $s$ by $t'^{\beta/\alpha}$ and using
the relation $W_{1}(x,t^{\beta/\alpha})=W(x,t)$ in Eq.(26), we
obtain
\begin{equation}
\label{(27)}
\begin{array}{lll}
\frac{\partial}{\partial t}W(x,t) &=&
\frac{\beta}{\Gamma(\alpha+1)}\frac{\partial}{\partial t}
\int_{0}^{t}t'^{\beta/\alpha-1}(t^{\beta/\alpha}-t'^{\beta/\alpha})^{\alpha-1}L_{FP}W(x,t')dt'\\
\\
&=& \Phi_{t}L_{FP}W(x,t),
\end{array}
\end{equation}
where
\begin{equation}
\label{(28)}
\Phi_{t}f(t)=\frac{\beta}{\Gamma(\alpha+1)}\frac{\partial}{\partial
t}
\int_{0}^{t}t'^{\beta/\alpha-1}(t^{\beta/\alpha}-t'^{\beta/\alpha})^{\alpha-1}f(t')dt'.
\end{equation}
When $\beta=\alpha$, Eq.(27) reduces into the time fractional FPE
(16). It is worthy to note that Mura et al. derived the integral
form of Eq.(27) from a non-Markovian forward drift equation and
called it stretched time-fractional diffusion equation [43].

The MSD $\langle x^{2}\rangle(t)$ of
$Y(t)=x(t^{\beta}Z^{-\alpha})$ in a free force field can be
computed by
\begin{equation}
\label{(29)}
\begin{array}{lll}
\frac{d}{dt}\langle x^2\rangle(t)
&=&\int_{-\infty}^{\infty}x^{2}\frac{\partial W(x,t)}{\partial t}dx\\
\\
&=&\frac{2\beta}{\Gamma(\alpha+1)}\frac{d}{dt}
\int_{0}^{t}t'^{\beta/\alpha-1}(t^{\beta/\alpha}-t'^{\beta/\alpha})^{\alpha-1}dt'\\
\\
&=&\frac{2\beta}{\Gamma(\alpha+1)}\frac{d}{dt}
\int_{0}^{t^{\beta/\alpha}}\frac{\alpha}{\beta}(t^{\beta/\alpha}-s)^{\alpha-1}ds\\
\\
&=& \frac{2\beta}{\Gamma(\alpha+1)}t^{\beta-1}.
\end{array}
\end{equation}
So, we have
\begin{equation}
\label{(30)} \langle
x^2\rangle(t)=\frac{2}{\Gamma(\alpha+1)}t^{\beta}, \hspace{0.5cm}
0<\alpha<1, 0<\beta<2.
\end{equation}

By comparing Eq.(30) with Eq.(18), we conclude that the diffusive
process $Y(t)$ is controlled by parameter $\beta$ in subordinator
$s(t)$. It is easy to obtain that the subordinated process $Y(t)$
is subdiffusive when $0<\beta<1$, normal diffusive when $\beta=1$
and supdiffusive when $1<\beta<2$.

\section{The subordinated process with subordinator $s(t)=t^{\beta}Z^{-\beta}$}

\ \ \ \ Here we assume that $x(\tau)$ satisfies Langevin equation
(6) and $s(t)$ has the form $s(t)=t^{\beta}Z^{-\beta} (0<\beta<2)$
where $Z$ is an $\alpha-$stable random variable $(0<\alpha<1)$.

Since
\begin{equation}
P(s(t)\leq \tau) =P(Z\geq \frac{t}{\tau^{1/\beta}})
=\int_{\frac{t}{\tau^{1/\beta}}}^{\infty}\psi_{\alpha}(t')dt',
\end{equation}
we have
\begin{equation}
\label{(32)}
g(\tau,t)=\frac{1}{\beta}\frac{t}{\tau^{1/\beta+1}}\psi_{\alpha}(\frac{t}{\tau^{1/\beta}}).
\end{equation}

After using the formulas (10) and (11), one can get the Laplace
transform of the PDF $g(\tau,t)$
\begin{equation}
\label{(33)}
\widetilde{g}(\tau,u)=\frac{\alpha}{\beta}\tau^{\alpha/\beta-1}u^{\alpha-1}\exp\{-u^{\alpha}\tau^{\alpha/\beta}\}.
\end{equation}

It is worth to note that the subordinated process $Y(t)$
constructed in this section can be connected to a CTRW with
correlated waiting times (see Ref.[21]) in terms of Eq.(33).
Indeed, in Ref.[21], the authors modeled CTRW with correlated
waiting times by the following coupled Langevin equations
\begin{equation}
\label{(34)} \frac{dx(\tau)}{d\tau}=\Gamma(\tau), \hspace{1cm}
\frac{dt(\tau)}{d\tau}=\int_{0}^{\tau}M(\tau-\tau')\eta(\tau')d\tau',
\end{equation}
where $M(\tau)$ is a non-negative continuous memory function and
$\eta(\tau)$ is a fully skew $\alpha-$stable L\'{e}vy noise. The
Laplace transform of the PDF $g(\tau,t)$ of the process $s(t)$
defined by Eq.(5) has the form
\begin{equation}
\label{35}
\widetilde{g}(\tau,u)=u^{\alpha-1}\phi'(\tau)\exp\{-u^{\alpha}\phi(\tau)\},
\end{equation}
where
$\phi(\tau)=\int_{0}^{\tau}d\tau'(\int_{\tau'}^{\tau}M(\tau''-\tau')d\tau'')^{\alpha}$.
By comparing Eq.(35) with Eq.(33), we have
$\phi(\tau)=\tau^{\alpha/\beta}$, which gives the connection
between the subordinated process $Y(t)$ constructed in this
section and the CTRW with correlated waiting times introduced in
Ref. [21].

Some other kinds of correlated CTRW processes were given in Refs.
[44-48]. The CTRW with correlated jumps was studied in Ref.[44].
The CTRWs with correlated waiting times defined as
$t(\tau)=\int_{0}^{\tau}|L_{\alpha}(\tau')|d\tau'$ in Refs.
[45-47] and with heterogeneous memorized waiting times defined as
$D_{0+}^{1-\mu}t_x(\tau)=|x|^{\theta/\gamma}|L_{\alpha}(\tau)|$ in
Ref. [48] are different from that introduced in Ref. [21].

Let us turn to the subordinated process $Y(t)$. In Laplace space,
the PDF $W(x,t)$ can be represented by
\begin{equation}
\label{(36)}
\widetilde{W}(x,u)=\int_{0}^{\infty}f(x,\tau)\widetilde{g}(\tau,u)d\tau.
\end{equation}

Inserting Eq.(33) into Eq.(36), we have
\begin{equation}
\label{(37)}
\begin{array}{lll}
\widetilde{W}(x,u)
&=& \int_{0}^{\infty}f(x,\tau)\frac{\alpha}{\beta}\tau^{\alpha/\beta-1}u^{\alpha-1}\exp\{-u^{\alpha}\tau^{\alpha/\beta}\}d\tau \\
\\
&=& u^{\alpha-1}\int_{0}^{\infty}f(x,s^{\beta/\alpha})e^{-u^{\alpha}s}ds \\
\\
&=& u^{\alpha-1}\int_{0}^{\infty}f_{1}(x,s)e^{-u^{\alpha}s}ds \\
\\
&=& u^{\alpha-1}\widetilde{f}_{1}(x,u^{\alpha}),
\end{array}
\end{equation}
where $f_{1}(x,\tau)=f(x,\tau^{\beta/\alpha})$. Obviously there
exists the relation
\begin{equation}
\label{(38)} \frac{\partial}{\partial
\tau}f_{1}(x,\tau)=\frac{\partial f(x,\tau_{1})}{\partial
\tau_{1}}\cdot \frac{d\tau_{1}}{d\tau},
\end{equation}
where $\tau_{1}=\tau^{\beta/\alpha}$.

Since the PDF $f(x,\tau)$ satisfies Eq.(7), we have
\begin{equation}
\label{(39)}
\begin{array}{lll}
\frac{\partial}{\partial \tau}f_{1}(x,\tau)
&=& \frac{\partial f(x,\tau_{1})}{\partial \tau_{1}}\cdot \frac{d\tau_{1}}{d\tau} \\
\\
&=& L_{FP}f(x,\tau_{1})\cdot\frac{d\tau_{1}}{d\tau} \\
\\
&=& L_{FP}f_{1}(x,\tau)\frac{\beta}{\alpha}\tau^{\beta/\alpha-1}.
\end{array}
\end{equation}

We Set $q(\tau)=\frac{\beta}{\alpha}\tau^{\beta/\alpha-1}$, whose
Laplace transform is
$\widetilde{q}(u)=\frac{\Gamma(\beta/\alpha+1)}{u^{\beta/\alpha}}$.
Taking Laplace transform for variable $\tau$ in Eq.(39), one gets
\begin{equation}
\label{(40)}
\begin{array}{lll}
u\widetilde{f}_{1}(x,u)-f_{1}(x,0)
&=&\int_{0}^{\infty}L_{FP}f_{1}(x,\tau)q(\tau)e^{-u\tau}d\tau\\
\\
&=&\frac{1}{2\pi
i}\int_{a-i\infty}^{a+i\infty}L_{FP}\widetilde{f}_{1}(x,s)\widetilde{q}(u-s)ds,
\end{array}
\end{equation}
with $0<a<Re(u)$.

From Eq.(37), with the initial condition
$W(x,0)=f_{1}(x,0)=\delta(x)$, we have
\begin{equation}
\label{(41)}
\begin{array}{lll}
u\widetilde{W}(x,u)-W(x,0)
&=& u^{\alpha}\widetilde{f}_{1}(x,u^{\alpha})-f_{1}(x,0)\\
\\
&=&\frac{1}{2\pi i}\int_{a-i\infty}^{a+i\infty}L_{FP}\widetilde{f}_{1}(x,s)\widetilde{q}(u^{\alpha}-s)ds\\
\\
&=& \frac{1}{2\pi
i}\int_{a-i\infty}^{a+i\infty}L_{FP}\widetilde{W}(x,s^{1/\alpha})s^{\frac{1-\alpha}{\alpha}}
\frac{\Gamma(\beta/\alpha+1)}{(u^{\alpha}-s)^{\beta/\alpha}}ds\\
\\
&=& \frac{1}{2\pi
i}\int_{a-i\infty}^{a+i\infty}s^{\frac{1-\alpha}{\alpha}}\frac{\Gamma(\beta/\alpha+1)}{(u^{\alpha}-s)^{\beta/\alpha}}ds
\int_{0}^{\infty}L_{FP}W(x,t')e^{-s^{1/\alpha}t'}dt'.
\end{array}
\end{equation}
Setting
$\widetilde{k}(u,s)=s^{\frac{1-\alpha}{\alpha}}\frac{\Gamma(\beta/\alpha+1)}{(u^{\alpha}-s)^{\beta/\alpha}}$
and taking inverse Laplace transform for $u\rightarrow t$ in
Eq.(41), we obtain
\begin{equation}
\label{(42)} \frac{\partial}{\partial t}W(x,t)
=\Phi_{t}L_{FP}W(x,t),
\end{equation}
where
\begin{equation}
\label{(43)} \Phi_{t}h(t)=\frac{1}{2\pi
i}\int_{a-i\infty}^{a+i\infty}k(t,s)ds
\int_{0}^{\infty}h(t')e^{-s^{1/\alpha}t'}dt'.
\end{equation}

Next we also compute the MSD $\langle x^{2}\rangle(t)$ of the
subordinated process $Y(t)=x(t^{\beta}Z^{-\beta})$ in a free force
filed.

Since
\begin{equation}
\label{(44)} \langle x^2\rangle(t) =2\int_{0}^{\infty}\tau
g(\tau,t)d\tau,
\end{equation}
we have
\begin{equation}
\label{(45)}
\begin{array}{lll}
\langle x^2\rangle(u)
&=& 2\int_{0}^{\infty}\tau \widetilde{g}(\tau,u)d\tau \\
\\
&=& 2\int_{0}^{\infty}\tau \frac{\alpha}{\beta}\tau^{\alpha/\beta-1}u^{\alpha-1}\exp\{-u^{\alpha}\tau^{\alpha/\beta}\}d\tau\\
\\
&=& 2u^{\alpha-1}\int_{0}^{\infty}s^{\beta/\alpha}e^{-u^{\alpha}s}ds\\
\\
&=& \frac{2\Gamma(\frac{\beta+\alpha}{\alpha})}{u^{\beta+1}}.
\end{array}
\end{equation}

So
\begin{equation}
\label{(46)} \langle
x^2\rangle(t)=\frac{2\Gamma(\beta/\alpha)}{\alpha\Gamma(\beta)}t^{\beta}.
\end{equation}

From Eq.(46), we conclude that the subordinated process is
subdiffusive when $0<\beta<1$, normal diffusive when $\beta=1$ and
supdiffusive when $1<\beta<2$.

\section{Conclusions}

\ \ \ \ In this work, we consider the subordinated process $Y(t)$
which is controlled by a family of subordinators. For the case of
$s(t)=t^{\alpha}Z^{-\alpha}$, the PDF $W(x,t)$ of the subordinated
process $Y(t)$ satisfies a time-fractional Fokker-Planck equation.
For the case of $s(t)=t^{\beta}Z^{-\alpha}$, the corresponding
diffusion equation is a stretched time-fractional Fokker-Planck
equation. For the case of $s(t)=t^{\beta}Z^{-\beta}$, we obtain
that the Laplace transform of the PDF $g(\tau,t)$ is of the form
$\widetilde{g}(\tau,u)=\frac{\alpha}{\beta}\tau^{\alpha/\beta-1}u^{\alpha-1}\exp\{-u^{\alpha}\tau^{\alpha/\beta}\}$.
The resulting process can be connected with CTRW with correlated
waiting time.

\section*{Acknowledgments}
This project was supported by the Natural Science Foundation of
China (Grant Nos. 11371016 and 11271311), the Chinese Program for
Changjiang Scholars and Innovative Research Team in University
(PCSIRT) (Grant No. IRT1179),
 the Lotus Scholars Program of Hunan province
of China.

\end{document}